\documentclass[12pt,fleqn]{article}
\usepackage{graphicx}
\usepackage{bm}

\def\shiftleft#1{#1\llap{#1\hskip 0.04em}}
\def\shiftdown#1{#1\llap{\lower.04ex\hbox{#1}}}
\def\thick#1{\shiftdown{\shiftleft{#1}}}
\def\b#1{\thick{\hbox{$#1$}}}

\begin{document}

\begin{center}
{\Large{\bf The two pion decay of the Roper resonance}}
\end{center}
\vspace{1cm}

\begin{center}
{\large{ E. Hern\'andez$^1$, E. Oset$^2$ and M.J. Vicente Vacas$^2$}}
\end{center}
\vspace{0.4cm}

\begin{center}
{\it $^1$ Grupo de F\'{\i}sica Nuclear}
\end{center}
\begin{center}
{\it Facultad de Ciencias, Universidad de Salamanca}
\end{center}
\begin{center}
{\it Plaza de la Merced s/n,}
\end{center}
\begin{center}
{\it 37008 Salamanca, Spain}       
\end{center}
\begin{center}
{\it $^2$Departamento de F\'{\i}sica Te\'orica e IFIC}
\end{center}
\begin{center} 
{\it Centro Mixto Universidad de Valencia-CSIC}
\end{center}
\begin{center}
{\it Institutos de Investigaci\'on de Paterna, Apdo. correos 22085,}
\end{center}
\begin{center}
{\it 46071 Valencia, Spain}       
\end{center}

\begin{abstract}

We evaluate the two pion decay of the Roper resonance in a model where
explicit
re-scattering of the two final pions is accounted for by the use of
unitarized chiral perturbation theory. Our model does not include  an explicit
 $\epsilon$ or $\sigma$  scalar-isoscalar meson decay mode, instead it generates it
dynamically
 by means of
the pion re-scattering.
The two ways, explicit or dynamically generated, of introducing this decay
 channel have very different amplitudes.
 Nevertheless, through
 interference
with the other terms of the model
we are able  to
 reproduce the same phenomenology as models with  explicit consideration
of the $\epsilon$ meson.

\end{abstract}

\section{Introduction}
  The Roper resonance is one of the controversial resonances, with an abnormally
large width  comparative to other resonances with larger mass.
 It appears naturally in quark models with a radial excitation of
one of the quarks \cite{Capstick:1994kb,Cano:1998wz,Bijker:1994yr}, but it 
has also been suggested that it
is dynamically generated by the meson-baryon interaction itself 
\cite{Krehl:2000km} and
thus would be essentially formed by a large meson-nucleon cloud.
  One of the intriguing properties of the Roper is its two pion decay mode.
  According to the  PDG \cite{Groom:2000in}, it has a 30-40 percent branching 
 ratio into
$N \pi \pi$, mostly going to $\Delta \pi$, and a small fraction of 5-10 percent
which goes into a nucleon and two pions in s-wave and isospin, I=0. This scalar
isoscalar mode plays a very important role in all reactions involving two pion
production close to threshold. The reason is that the contribution from the
nucleon intermediate states cancels at threshold when the direct and crossed
terms are taken into account. Then the next resonance which is the $\Delta$
involves p-wave couplings which would vanish at threshold, and finally comes the
Roper, which thanks to this non vanishing scalar isoscalar decay mode into two
pions gives a non vanishing contribution to the threshold amplitudes. This has
been shown explicitly to be the case in pion induced two pion production
\cite{Oset:1985wt,Sossi:1993zw,Bernard:1995gx,Jensen:1997em}, photon induced 
two pion production \cite{GomezTejedor:1994bq,GomezTejedor:1996pe}
and two pion production in nucleon nucleon collisions 
\cite{Alvarez-Ruso:1998mx,Alvarez-Ruso:1998xg,Alvarez-Ruso:2000jq,bo}.
The influence of the Roper excitation and its decay modes in many other
reactions has been discussed in
\cite{Soyeur:1999jz, Soyeur:1998iu, Soyeur:2000fm}.
Similarly, it has also been shown \cite{Morsch:1992vj,Hirenzaki:1996js} 
in the study of the Roper
excitation in the ($\alpha , \alpha '$) reaction that the Roper is very 
efficiently
excited by an isoscalar source, which should somehow be related to this scalar
isoscalar decay.

  The evidence for this scalar isoscalar two pion decay comes mostly from the
analysis of Manley \cite{Manley:1984jz,Manley:1992yb}, where he fits the data by means of the
excitation of an $\epsilon$ meson of about 800 MeV mass and width plus the decay
into $\Delta \pi$. This
$\epsilon$ meson is what we would  call now the $\sigma$ meson, or in an
alternative nomenclature, the $f_0(400-900)$ meson, as also used in the PDG.
However, the immediate conceptual problem arises, since, whatever this meson is
called, the strength of the two pion distribution does not follow the shape
provided by the exchange of such a heavy and broad meson. The isoscalar 
two pion distribution at low energies is governed by the scattering matrix of 
two pions in L=0, I=0,
which has a broad bump consisting of a large background on top of the effects of
the $\sigma$ pole, which is now present in all modern theoretical
\cite{Oller:1997ti,OllOsePel,Colangelo:2001df} and experimental analyses
\cite{Ishida:2001pt,Kaminski:1999ns}, (see
also the proceedings of the $\sigma$ Workshop \cite{kyoto} ).  The pole of the
$\sigma$ appears in all these approaches with a mass around 500 MeV and a width
around 400 MeV, and, as mentioned above, there is also a large  background present
in the $\pi \pi$ t-matrix. 

    In the present context it is also worth mentioning that the use of chiral
perturbation theory \cite{Gasser:1985ux} and its unitary extensions in coupled channels
\cite{Oller:1997ti,OllOsePel,Oller:1999zr,Kaiser:1998fi,Nieves:1999hp,
Nieves:2000bx,Markushin:2000fa} 
has brought a new perspective on the
nature of the scalar meson resonances, in particular the $\sigma$.  Indeed, what
is found in these works is that the $\sigma$ is generated dynamically from the
lowest order chiral Lagrangian and the multiple scattering of the mesons 
implicit in the unitary approach. This finding and the previous statement is
more than semantics, because it implies that anything having to do with the
production of a $\sigma$ should be considered as the production of two pions
which undergo final state interaction (in this case in the strong L=0, I=0
channel).  This is the
philosophy taken in the present work, where we perform a theoretical study 
of the two
pion decay of the Roper, with the explicit consideration of the final state
interaction of the two pions, which automatically generates the production of
this two pion $\sigma$ mode.  The approach is hence different to the one
followed in \cite{Manley:1984jz,Manley:1992yb} since we do not allow the direct 
production of a
$\sigma$. Yet, we aim at reproducing the same phenomenology that was fitted in
the analysis of \cite{Manley:1984jz,Manley:1992yb}.  We will show how this is 
possible, and even if a
very different, and obviously more realistic, distribution of the two pion 
invariant mass is obtained for the scalar isoscalar decay mode, the coherent
sum of the different mechanisms that we have leads to mass distributions of the
two pions or of one pion-nucleon system, in agreement with the
results that one would obtain with Manley's approach of $\Delta \pi$  plus
the massive and broad $\epsilon$ meson production.

\section{Manley's approach for the two pion decay of the Roper}
In this section we will translate the two pion decay model of the Roper described in
refs. \cite{Manley:1984jz,Manley:1992yb} to the language of Lagrangian
and Feynman 
diagrams. 
As explained in the introduction, the model has two intermediate decay channels for
the Roper decay into two pions: 
the  Roper  can decay either
into an intermediate $\Delta \pi$ state or into an intermediate $N \epsilon$ state. The 
$\epsilon$ meson  is a scalar isoscalar meson
with a mass and width of $M_{\epsilon}\!=\!
\Gamma_{\epsilon}(M_{\epsilon})\!=\!800$ MeV. The Feynman diagrams for the two decaying modes appear
in Fig. 1a-b.
\begin{figure}[h]
\begin{center}
\resizebox{10cm}{6cm}{\includegraphics{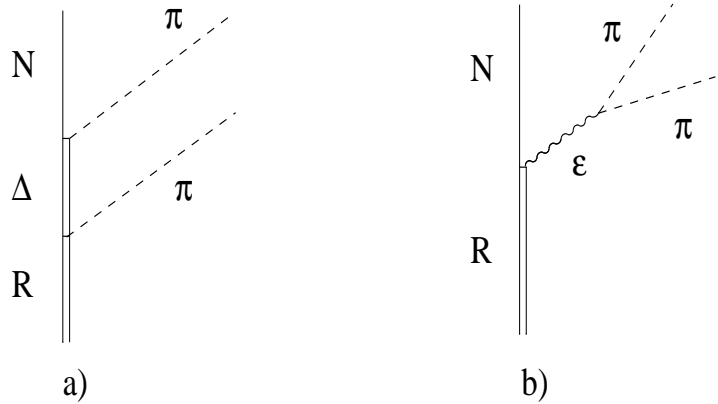}}
\parbox[b]{13cm}{\caption{Feynman diagrams for the two processes contributing to the
Roper decay into two pions in Manley's approach. a) $\Delta\pi$ intermediate channel. 
b) $N\epsilon$ intermediate channel.}}
\end{center}
\end{figure}
\noindent
In the spirit of refs. \cite{Manley:1984jz,Manley:1992yb} we will use here
non relativistic Lagrangian  for the different vertices involved in the calculation. Those
Lagrangian are given, with obvious notation, as
\begin{eqnarray}
\label{la1}
{\cal L}_{R\Delta\pi}(x)&=& \frac{f_{R\Delta\pi}}{m_{\pi}}\Psi^{\dagger}_{\Delta}(x)
S^{\dagger}_j {\bf T}^{\dagger}(\partial_j \b{\pi}(x))\Psi_R(x)
\,\epsilon(x) \nonumber\\
{\cal L}_{\Delta N\pi}(x)&=& \frac{f_{\Delta N \pi}}{m_{\pi}}\Psi^{\dagger}_{N}(x)
S_j {\bf T}(\partial_j \b{\pi}(x))\Psi_{\Delta}(x)
\end{eqnarray}
and 
\begin{eqnarray}
\label{la2}
{\cal L}_{RN\epsilon}(x)&=& g_{RN\epsilon}\Psi^{\dagger}_N(x)\Psi_R(x)\,\epsilon(x) \nonumber\\
{\cal L}_{\epsilon\pi\pi}(x)&=& g_{\epsilon\pi\pi}\
\epsilon(x)\,\b{\pi}(x)\,\b{\pi}(x)
\end{eqnarray}
${\bf S}^{\dagger}$ and ${\bf T}^{\dagger}$ are respectively the spin and isospin operators for the 1/2 to 3/2 transition with matrix
elements just given by Clebsch-Gordan coefficients.

The amplitude for the process in Fig. 1a is given by
\begin{eqnarray}
\label{a1a}
\hspace{-1cm}{\cal M}_{1a}\hspace{-.2cm}&=&\hspace{-.2cm} i\,
\frac{f_{R\Delta\pi}}{m_{\pi}}\,
\frac{f_{\Delta N \pi}}{m_{\pi}}\
\Phi_N^{\dagger}\chi_N^{\dagger} \nonumber\\
&&\hspace{-.4cm}\bigg[\
\hspace{.4cm}\bigg(
\frac{2}{3}\,{\bf q}_1\cdot{\bf q}_2-\frac{i}{3}\,\b{\sigma}\cdot
({\bf q}_1\times{\bf q}_2)
\bigg)
\
\bigg(
\frac{2}{3}\,\delta_{j_1 j_2}-\frac{i}{3}\,\varepsilon_{j_1 j_2 k}
\cdot\tau_k
\bigg)\nonumber\\
&&\hspace{.4cm}\frac{M_{\Delta}}{E_{\Delta}({\bf q}_1)}\
\frac{1}{M_R-q_1^0-E_{\Delta}({\bf q}_1)
+\frac{i}{2}\,
\Gamma_{\Delta}(p_R-q_1)}\nonumber\\
&& + (1\leftrightarrow 2)\hspace{.4cm}
\bigg]\,\Phi_R\chi_R
\end{eqnarray}
$\Phi$ and $\chi$ are respectively isospin and spin Pauli spinors for both
the nucleon and  the Roper while $\b{\tau}$ and $\b{\sigma}$  are the
isospin and spin Pauli matrices.  $p_R=(M_R,{\bf 0})$, $q_1$ and $q_2$ are the
four-momentum of the
Roper and the two pions, and
$E_{\Delta}({\bf q})$ is the on-shell energy of a $\Delta$ with
three-momentum ${\bf q}$. Finally $j_1$ and $j_2$ are isospin indices
in the cartesian base
for the two pions. Note that since we are working in the isospin mathematical base
the pions are identical particles and the amplitude has to be symmetrized with respect to
the exchange of the pions. In the expression for the width we have to include a factor
$1/2$ of symmetry.
For the process in Fig. 1b we get 
\begin{eqnarray}
\hspace{-1cm}{\cal M}_{1b}= -2i\,g_{RN\epsilon}\, g_{\epsilon\pi\pi}
\ \frac{1}{s-M_{\epsilon}^2+i\,M_{\epsilon}\,\Gamma_{\epsilon}(s)}
\ \Phi_N^{\dagger}\cdot\Phi_R\ \ \chi_N^{\dagger}\cdot\chi_R\hspace{.3cm}
\delta_{j_1 j_2}
\end{eqnarray}
where  $s$ is the total four-momentum square of the two final pions.
 Even though the previous amplitudes are
non-relativistic for  the phase space integrals we shall take into account all
the appropriate relativistic factors.
As for the coupling constants we fix $f_{\Delta N \pi}$ from the width of the $\Delta$
as given by the PDG. A value of $f_{\Delta N \pi}\!=\!2.07$ results. For the other coupling
constants one needs information on the width of the Roper associated to the two channels
involved. Those values and the  mass of the Roper we take from
Manley's analysis. Comparing $\Gamma_{R\to \Delta \pi}\!=\!88$ MeV to the result obtained
with the use of the Feynman diagram
of Fig. 1a one gets 
\hbox{$f_{R\Delta \pi}\!=\!1.56$}, while from $\Gamma_{R\to N \epsilon}\!=\! 33$ MeV and 
the evaluation of
the contribution due to the Feynman diagram in Fig. 1b one determines the product
of coupling constants \hbox{$g_{RN\epsilon}\cdot g_{\epsilon\pi\pi}\!=\!1.43\cdot 
10^{-2}$\,$M_{\epsilon}^2$\,MeV$^{-1}$}.
The  sign of this product of coupling constants relative to
$f_{R\Delta\pi}$ is chosen to be positive in accordance with the sign assignment
done in \cite{Manley:1992yb}.
Note also that for the first process
the pions can
be in both $I=0,1$ isospin states while for the second only $I=0$
is allowed.
The total width for the Roper decay into two pions that we 
 get when taking into account both terms is given by
$\Gamma_{R\to N\pi \pi}\!=\!153$ MeV, meaning that there is constructive interference
between the two contributions. While this interference effect is also present in the
amplitude used in ref.  \cite{Manley:1992yb} there the total width is taken to be
$\Gamma_{R\to N\pi \pi}\!=\Gamma_{R\to \Delta \pi}+\Gamma_{R\to N \epsilon}=121$ MeV.

\begin{figure}[h!]
\begin{center}
\rotatebox{270}{
\resizebox{8cm}{13cm}{
\includegraphics{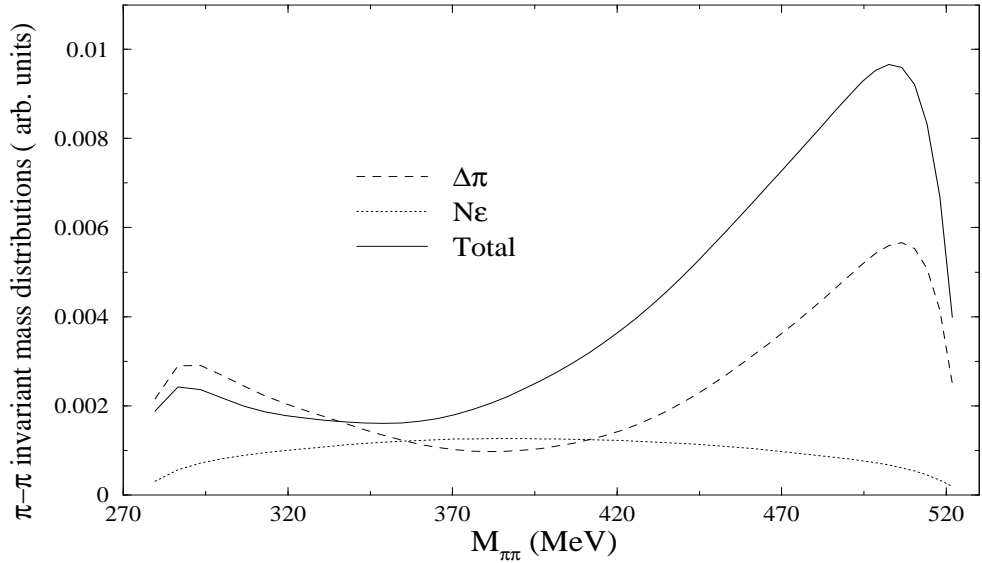}
}
}
\parbox[b]{13cm}{\caption{$\pi-\pi$ invariant mass distributions as obtained in Manley's approach.
Dashed line: distribution obtained from considering the $\Delta\pi$ intermediate channel
alone. Dotted line: distribution obtained from considering the $N\epsilon$
intermediate channel alone. Full curve: total result.}}
\end{center}
\end{figure}

In Fig. 2 we give  two-pion invariant mass distributions obtained within the
approach
just described. The distribution that arises from
 considering the 
intermediate channel $N\epsilon$ alone is represented by the dotted line
and it is essentially given by phase space. The dashed line gives now the distribution 
corresponding to
 the  intermediate $\Delta\pi$
state. There one sees two peaks at large and small invariant masses
that are due to the presence of a $(\vec{q}_1\cdot\vec{q}_2)^2$ term,
with $\vec{q}_{1,2}$ the three-momentum of the pions, 
in the amplitude square (see Eq. (\ref{a1a})). The maximum 
for this quantity is reached when the two pions move in the same direction (small 
invariant mass)
or in opposite direction (large invariant mass). The invariant mass distribution 
corresponding
to the coherent sum of the two channels is given by the full curve. The interference effect
is clearly seen at large and small invariant masses. While the peak at large invariant mass
increases the one at low invariant mass decreases.

In Fig. 3 we show  pion-nucleon invariant mass distributions. As before the $N\epsilon$
contribution is basically given by phase space, while the $\Delta\pi$ one shows a peak below
the Delta mass. In the total contribution  the central peak is enhanced 
due to interference.
\begin{figure}[h!]
\begin{center}
\rotatebox{270}{
\resizebox{8cm}{13cm}{
\includegraphics{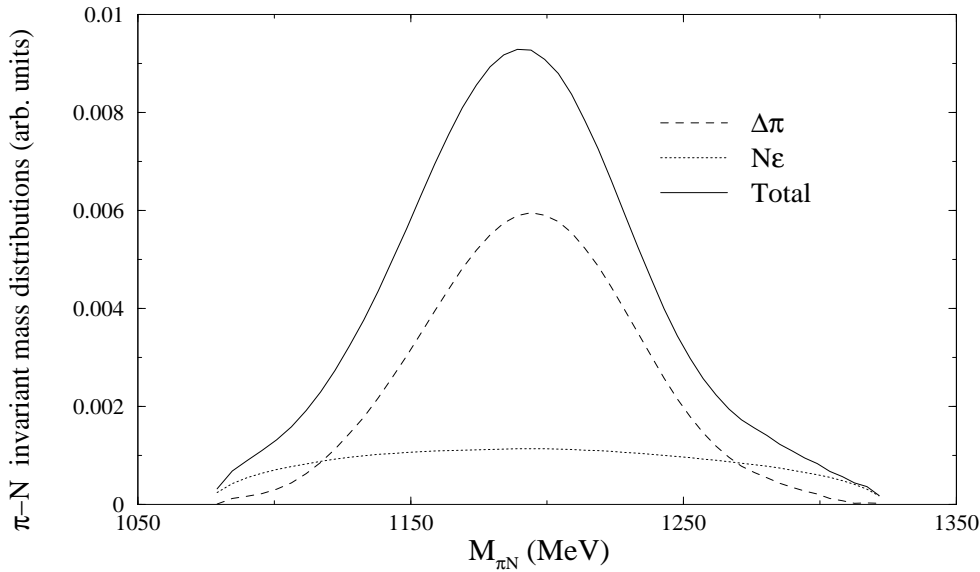}
}
}
\parbox[b]{13cm}{\caption{$\pi-N$ invariant mass distributions as obtained in Manley's approach.
Notation as in Fig. 2.}}
\end{center}
\end{figure}

As these invariant mass distributions are obtained from amplitudes that are fitted to
experiment we will consider them as if they were true ``experimental results'' to which
we can compare our own results.  

\section{Model for the two pion decay of the Roper}
In our model we will consider two types of contributions. In the first type, that we shall
call open-diagrams contribution,
 the Roper decays into a baryon and a pion and then the baryon  decays into
nucleon-pion. No final state interaction between the two pions is considered. For 
the intermediate baryon we take Delta, nucleon and Roper itself.
\begin{figure}[h]
\begin{center}
\resizebox{10cm}{5cm}{\includegraphics{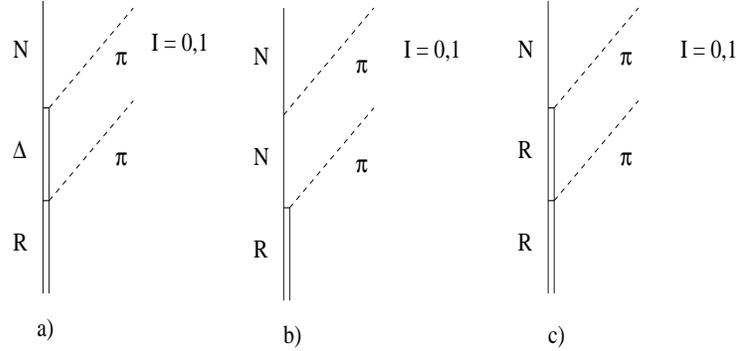}}
\parbox[b]{13cm}{\caption{
Feynman diagrams corresponding to our ``open-diagrams'' contribution
to the Roper decay into two pions. Diagrams a), b) and c) correspond to
taking Delta, nucleon or Roper as intermediate baryons. The two final
 pions can be in both $I=0,1$ isospin
 states.}}
\end{center}
\end{figure}

This contribution, whose Feynman diagrams  appear in Fig. 4,  is
the equivalent in our model of the $\Delta\pi$ intermediate channel in Manley's approach.
The differences come from the fact that we include also $N\pi$ and $R\pi$ intermediate
states and we take into account relativistic
corrections for the vertices involved. Those relativistic corrections are given 
for the $N'\to N\pi$ transition as
\begin{eqnarray}
\b{\sigma}\cdot{\bf q} \to \b{\sigma}\cdot{\bf q}\ (1-\frac{q^0}{2M'})
-\b{\sigma}\cdot{\bf p}\ (\frac{q^0}{2M'}+\frac{q^0}{2M})
\end{eqnarray}
where $N',N$ stand for either nucleon or Roper,
$(q^0,{\bf q})$ is the pion four momentum and ${\bf p}$ the three-momentun 
of the final $N$. This comes automatically from the evaluation of the matrix
elements of the $\gamma^\mu \gamma_5 \partial_\mu$ operator of the pseudovector
coupling of the pions to the nucleons.
For the vertices involving $\Delta$ we take the Lagrangian of Eq. \ref{la1}, 
where it has been implicitly assumed the $\Delta$ to be at rest, and introduce the
appropriate modification
\begin{eqnarray}
&&{\bf S}^{\dagger}\cdot{\bf q} \to {\bf S}^{\dagger}\cdot( {\bf q}-\frac{q^0}{M_{\Delta}}
\,{\bf p}_{\Delta})\nonumber \\
&&{\bf S}\cdot{\bf q} \to {\bf S}\cdot( {\bf q}-\frac{q^0}{M_{\Delta}}
\,{\bf p}_{\Delta})
\end{eqnarray}
For the coupling constants we take $f_{NN\pi}=g_A\frac{m_{\pi}}{2f_{\pi}}=0.95$, the
naive quark model result
$f_{RR\pi}=f_{NN\pi}$, and for $f_{RN\pi}$ we use the experimental information on
$\Gamma_{R\to N\pi}=270$ MeV from which  $f_{RN\pi}=0.40$. We still have to fix
$f_{R\Delta\pi}$.

For that we  need our second type of contribution, that we call closed-diagrams contribution,
and that we construct by allowing the two 
pions in the final state to re-scatter in the $I=0$ isospin channel.
The corresponding Feynman diagrams   
appear in Fig. 5 and this is our model analogue
of the $N\epsilon$ channel in Manley's analysis.
\begin{figure}[h]
\begin{center}
\resizebox{10cm}{5cm}{\includegraphics{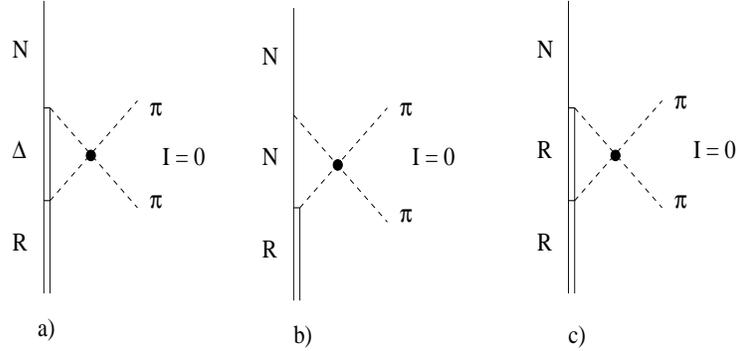}}
\parbox[b]{13cm}{\caption{
Feynman diagrams corresponding to our ``closed-diagrams'' contribution
to the Roper decay into two pions. a), b) and c) as in Fig. 4. The big dot represents 
the $\pi\pi$ t-matrix in the $I=0$ channel. Re-scattering in the $I=1$ channel is neglected.  
 }}
\end{center}
\end{figure}

To fully define our model we have to
give the form factors that we use to regularize the loop integrals  and also the $\pi\pi$
 t-matrix for the channel \hbox{$L=0,I=0$}. The latter will
be given in the next subsection. 
For each of the baryon-baryon-pion vertices we take a monopole form factor
\begin{equation}
F({\bf q})=\frac{\Lambda^2}{\Lambda^2+{\bf q}^2}
\end{equation}
 with the same value of
$\Lambda$ for all cases. We will discuss the cut-off dependence in the
results section.

As an illustration we give the amplitude corresponding to the process in
Fig. 5a
\begin{eqnarray}
\hspace{-2cm}{\cal M}_{5a}\hspace{-.2cm}&=&
\hspace{-.2cm} \frac{2}{3\sqrt3}\ 
\frac{f_{R\Delta\pi}}{m_{\pi}}\,
\frac{f_{\Delta N \pi}}{m_{\pi}}\ \Phi_N^{\dagger}\ \chi_N^{\dagger}\
\bigg\{ \int \frac{d^4q}{(2\pi)^4}\nonumber\\
&&\hspace{-.4cm}\bigg\{ \
\bigg[\
\ \bigg(
\frac{2}{3}\,({\bf p}_N-{\bf q})\cdot{\bf q}-\frac{i}{3}\,\b{\sigma}\cdot
(({\bf p}_N-{\bf q})\times{\bf q}) 
\bigg)
\bigg(
1+\frac{M_R}{M_{\Delta}}-\frac{q^0}{M_{\Delta}}
\bigg)
\nonumber\\
&&+\ \frac{2}{3}\ {\bf q}^2\ \frac{q^0-p_N^0}{M_{\Delta}}\ \bigg]\ 
\frac{M_{\Delta}}{E_{\Delta}({\bf  q})}\
\frac{1}{q^0-E_{\Delta}({\bf q})
+\frac{i}{2}\,
\Gamma_{\Delta}(q)}\nonumber\\
&&\frac{1}{(p_N^0-q^0)^2-({\bf p}_N-{\bf q})^2-m^2_{\pi}+i\varepsilon}\
\nonumber\\
&&\frac{1}{(M_R-q^0)^2-{\bf q}^2-m^2_{\pi}+i\varepsilon}
\ \frac{\Lambda^2}{\Lambda^2+({\bf p}_N-{\bf q})^2}
\ \frac{\Lambda^2}{\Lambda^2+{\bf q}^2}\ \bigg\}
\nonumber\\                            
&& t^{I=0}_{\pi\pi}(s)\ \
 \bigg\}\ \Phi_R\ \chi_R
\end{eqnarray}
Here $q$ and $p_N$ are respectively the four-momentum of the $\Delta$
in the loop  and
 of the final nucleon. For the two
final pions we have already taken into account the fact that they are in
an $I=0$ isospin state. Similar results are obtained for the case of
intermediate nucleon (Fig. 5b) or Roper (Fig. 5c).
All baryon propagators that we use throughout the paper contain a momentum dependent imaginary part. 
In all cases we use the expression for the width corresponding to the $N\pi$ decay channel. 
For the Roper
we re-scale that result to its total width. 

We fix the $f_{R\Delta\pi}$ coupling constant by fitting with our full model,
including the $\pi\pi$ t-matrix given in the next subsection, 
the total Roper decay width into two pions. 
For this width we take $\Gamma_{R\to N\pi\pi}=153$ MeV as evaluated in the previous
section and obtain a coupling constant that depends on the cut-off $\Lambda$. 
The best agreement with the invariant mass distributions is obtained with 
$\Lambda=0.7$ GeV which gives  $f_{R\Delta\pi}=1.1$ to be compared
to 1.56 obtained in Manley's approach. Larger $\Lambda$ values as those used in
ref. \cite{Oset:2000} also agree reasonably with data.


 
\subsection{$\pi\pi$ t-matrix in the $L=0,I=0$ channel}
The $\pi\pi$ t-matrix in the $L=0,I=0$ channel we take from refs.
\cite{Oller:1997ti,OllOsePel} where they use a non-perturbative approach
 that combines
pion-pion potentials provided by the lowest order chiral Lagrangian with the
Lipmann-Schwinger 
equation. The pion-pion potential corresponding to Fig. 6 is given by
\begin{equation}
V_{\pi\pi}^{I=0}=-\frac{6}{f_{\pi}^2}\bigg(
s-\frac{m_{\pi}^2}{2}-\frac{1}{3} \sum_j (q_j^2-m_{\pi}^2)\bigg)
\end{equation}
where $f_{\pi}=92.4$ MeV is the pion decay constant and $\sqrt{s}$ gives the total energy
in the center of mass.
We give the potential separated into on-shell plus off-shell parts.
\begin{figure}[h]
\begin{center}
\resizebox{5cm}{3cm}{\includegraphics{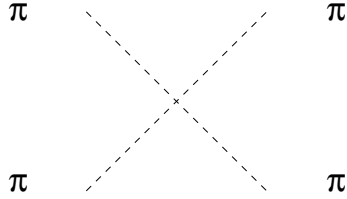}}
\parbox[b]{13cm}{\caption{
Feynman diagram for the $\pi\pi$ potential.
 }}
\end{center}
\end{figure}

To obtain the t-matrix  one sums an infinite set of diagrams where pions are allowed
to re-scatter in the s-channel. Formally on can write this Lipmann-Schwinger equation as
\begin{equation}
\label{ls}
t_{\pi\pi}^{I=0}=V_{\pi\pi}^{I=0}+V_{\pi\pi}^{I=0}\cdot G \cdot t_{\pi\pi}^{I=0}
\end{equation}
where $G$ stands for the loop in Fig. 7.
\begin{figure}[h]
\begin{center}
\resizebox{13cm}{3cm}{\includegraphics{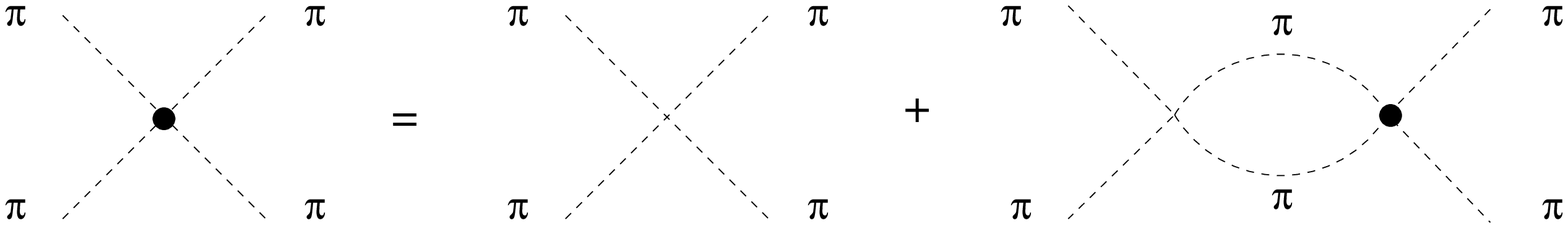}}
\parbox[b]{13cm}{\caption{
Set of Feynman diagrams corresponding to the
Lipmann-Schwinger equation used to evaluate
the $\pi\pi$ t-matrix in the $I=0$ channel. The contribution of kaons in the
loop is neglected.
 }}
\end{center}
\end{figure}

As shown in ref. \cite{Oller:1997ti}
the use of the off-shell part of the potential amounts to a renormalization of $f_{\pi}$
and the pion mass and then it should not be included if one uses the physical
values for those quantities. In that case Eq.(\ref{ls}) is  purely
algebraic and the factor $G$ is given by
\begin{equation}
G(s=P^2)=\frac{i}{6}\int \frac{d^4q}{(2\pi)^4}\ \frac{1}{q^2-m_{\pi}^2+
i\varepsilon}\
\frac{1}{(P-q)^2-m_{\pi}^2+i\varepsilon}
\end{equation}
where $P$ is the total four-momentum of the two incoming (outgoing) pions.
The integral is divergent and has to be regularized. Here we follow ref. \cite{OllOsePel}
where dimensional regularization is used. The regularization mass, that is
treated in ref. \cite{OllOsePel} as a free parameter, is given by $\mu=1.2$ GeV. The final expression for
$G(s)$ is
\begin{equation}
G(s)=\frac{1}{6\,(4\pi)^2}\bigg(-1+\ln \frac{m_{\pi}^2}{\mu^2}
+\sigma \ln \frac{\sigma+1}{\sigma-1}-i\pi\sigma\bigg)
\end{equation}
where $\sigma=\sqrt{1-4m_{\pi}^2/s}$, and then
\begin{equation}
t_{\pi\pi}^{I=0}=-\frac{6}{f_{\pi}^2}\frac{s-m_{\pi}^2/2}
{1+\frac{1}{f_{\pi}^2}(s-m_{\pi}^2/2) G(s)}
\end{equation}

We shall use this on-shell t-matrix even though the two pions in the loop in Fig. 7
can be off-shell. It has been shown in ref. \cite{Oset:2000} that for the case of
only nucleons and Deltas involved the dominant off-shell contribution is 
canceled
exactly by considering, at the same order in the chiral counting, diagrams that contain
vertices with one baryon line and three pions. By analogy a three pion vertex
involving the Roper should also be included to produce this cancellation in
the Roper case. Hence, as in \cite{Oset:2000}, we shall only take the on shell
part of the $\pi \pi$ interaction.

\section{Results and discussion}
In Table 1 we show the contribution to the
Roper decay width into two pions of our open- and closed-diagrams mechanisms 
for $\Lambda=0.7$ GeV. Unless otherwise indicated all results correspond to this
cut-off value.
The results that we obtain are compared
to the contributions of the equivalent $\Delta\pi$ and $N\epsilon$ mechanisms in 
Manley's approach. 
The role of the
mechanisms 
is different in the two models.
\begin{table}[h!]
\label{table:tab1}
\begin{center}
\nobreak
\begin{tabular}[t]{ |lr|  lr|  } 
\hline
& & &\\
 \hspace{1cm} Our model & & \hspace{1cm}Manley's approach &\\ 
& & &\\
\hline 
 & & & \\ 
$\bullet$\ Open-diagrams & 55.3 &$\bullet$\   $\Delta\pi$ & 88 \\  
& & &\\
\ \ - Delta alone & 54.7 & &\\
\ \ - No Delta    &  0.6 & &\\
 & & & \\
\hline
& & &\\
$\bullet$\ Closed-diagrams & 68.6  &$\bullet$\  $N\epsilon$ & 33\\  
& & &\\
\ \ - Delta alone & 31.8 & &\\
\ \ - No Delta    &  8.2 & &\\
 & & & \\
\hline
& & &\\
$\bullet$\ Coherent sum & 153 &$\bullet$\  Coherent sum& 153\\
 & & &\\                    
\hline 
\end{tabular}
\parbox[b]{14cm}{\caption{ Contribution of the different mechanisms to the two-pion Roper decay width.
All numbers are in MeV. ``Delta alone'' in  our model results  
means that we consider 
the Delta as the only intermediate baryon. ``No Delta'' in  
our model results means 
that we take only nucleon and Roper as intermediate baryons. Note the
constructive
interference between the ``Delta alone'' and ``No Delta'' contributions in the Closed-diagrams case.}}

\end{center}
\end{table}
While in Manley's approach the dominant contribution comes from the $\Delta\pi$ mechanism
in our case the closed-diagrams contribution, whose analogue in Manley's
analysis is the $N\epsilon$ channel, is larger.
Also from the numbers in Table 1 one sees that in our model
the largest contribution to the width comes from considering the 
Delta to be the intermediate baryon. The contribution that comes from considering intermediate 
nucleon and Roper is very small and only for the case of
closed-diagrams the interference with the dominant intermediate Delta contribution is of
some relevance.
\begin{figure}[h!]
\begin{center}
\rotatebox{270}{
\resizebox{8cm}{13cm}{
\includegraphics{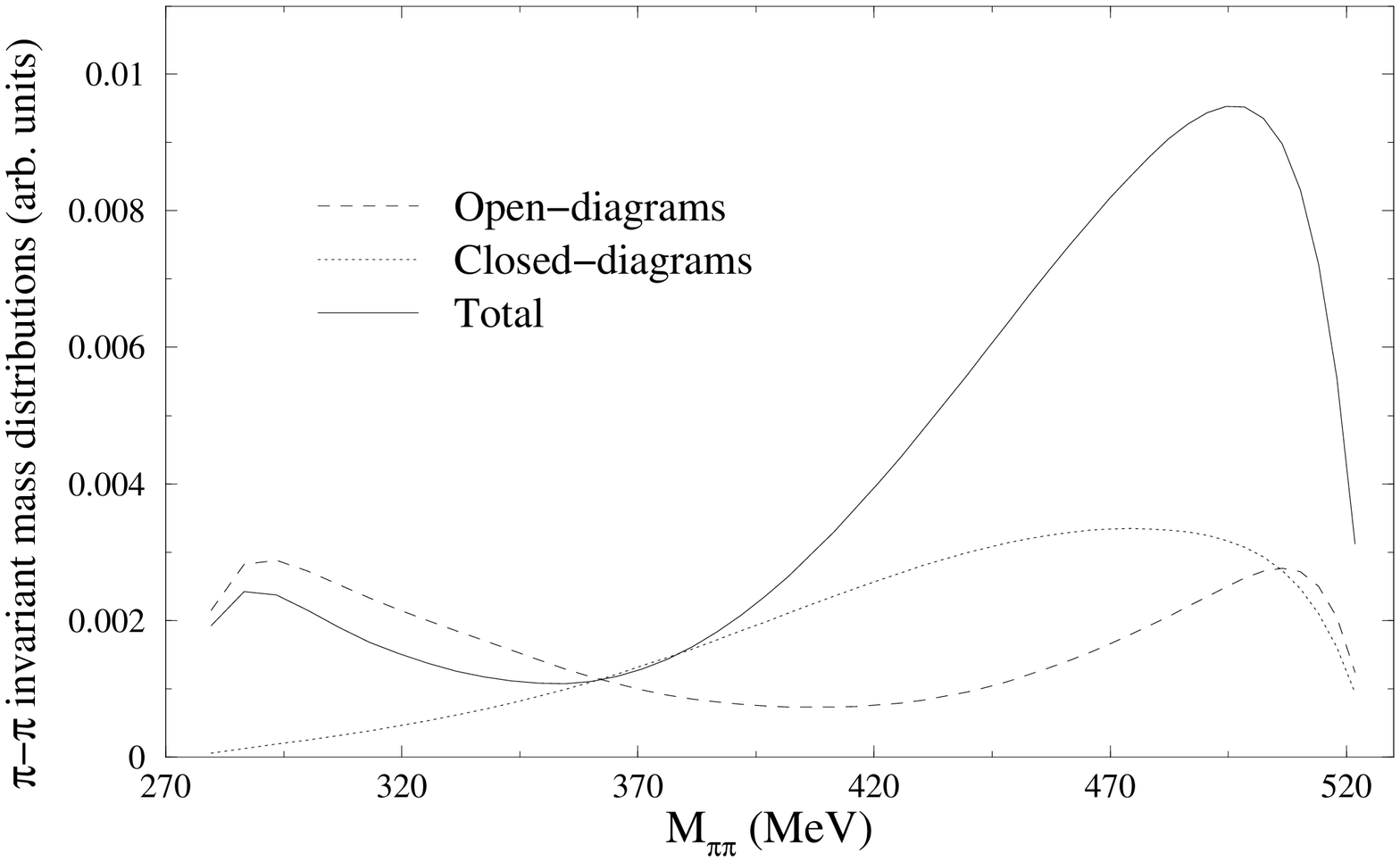}
}
}
\parbox[b]{14cm}{\caption{
$\pi-\pi$ invariant mass distributions as obtained in our model.
Dashed line: distribution obtained from considering the open-diagrams contribution alone.
 Dotted line: distribution obtained from considering the closed-diagrams
contribution alone. Full curve: total result.
}}
\end{center}
\end{figure}

In Fig. 8 we show now the two-pion invariant mass distribution obtained in our model.
The dashed line corresponds to the open-diagrams contribution. One sees the two peak
structure associated to the presence of the Delta. Compared to the $\Delta\pi$ distribution
in Manley's approach we see that the peak at high invariant mass is
reduced in our case. This is due to the relativistic corrections present in our model.
The dotted line corresponds to the closed-diagrams contribution. It has a broad peak
at high invariant mass while it goes to zero at low invariant mass. Its shape is
totally different from the phase space shape of the corresponding $N\epsilon$ contribution
in Manley's analysis. These differences notwithstanding, we see that the distribution that
corresponds to the coherent sum of the two channels resembles very much what
one gets in Manley's approach.
The two total distributions are compared now in Fig. 9 where
one sees a good agreement between the two different approaches. Also in Fig. 9 we show
the results for other values of the cut-off. This gives an idea of the theoretical uncertainties
in our calculation. 
\begin{figure}[h!]
\begin{center}
\rotatebox{270}{
\resizebox{8cm}{13cm}{
\includegraphics{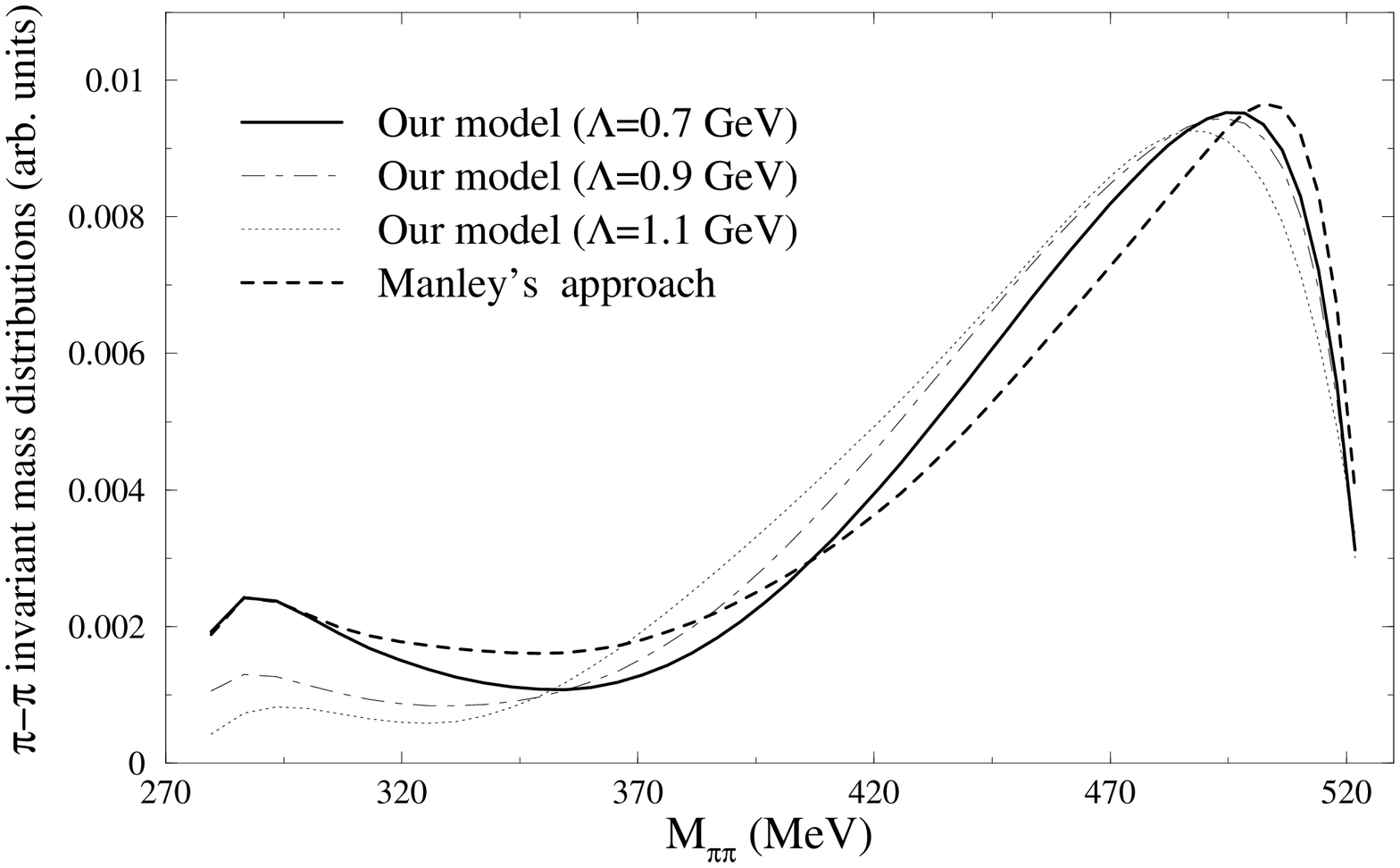}
}
}
\parbox[b]{14cm}{\caption{
Comparison of the $\pi-\pi$ invariant mass distributions as obtained 
in the two models. Dashed line: Manley's approach distribution. 
Our model result: with $\Lambda=0.7$\,GeV $(f_{R\Delta\pi}=1.1$) solid line,  
$\Lambda=0.9$\,GeV ($f_{R\Delta\pi}=0.83$)
dashed-dotted line, $\Lambda=1.1$\,GeV  ($f_{R\Delta\pi}=0.62$)  dotted line.    
}}
\end{center}
\end{figure}

A similar result is obtained for the pion-nucleon invariant mass distributions depicted in
Fig. 10.
\begin{figure}[h!]
\begin{center}
\rotatebox{270}{
\resizebox{8cm}{13cm}{
\includegraphics{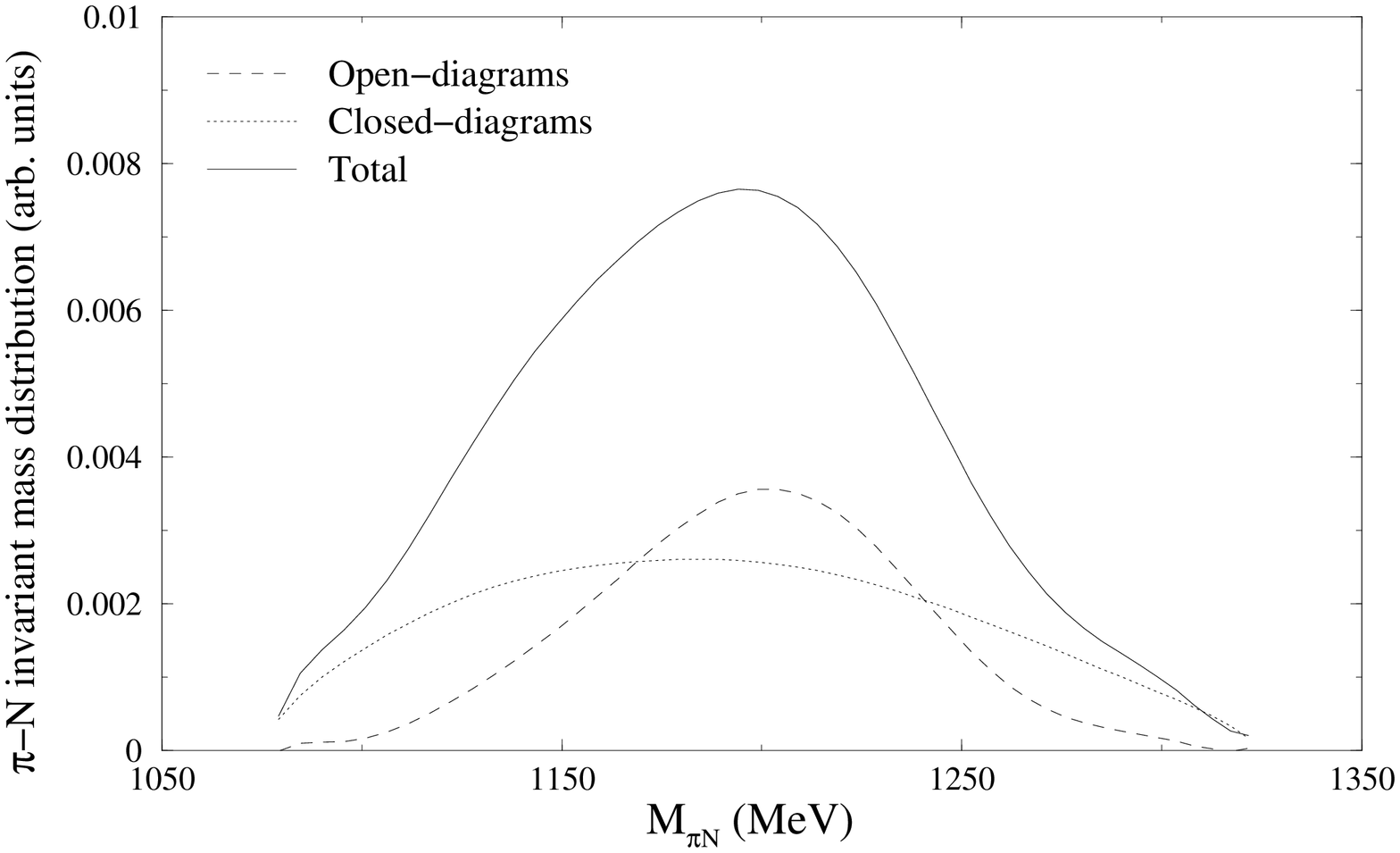}
}
}
\parbox[b]{14cm}{\caption{
$\pi-N$ invariant mass distributions as obtained in our model.
Notation as in Fig. 8
}}
\end{center}
\end{figure}
The dashed line gives again the distribution corresponding to our open-diagrams
mechanism. A peak is clearly seen around the Delta mass. The distribution corresponding
to our closed-diagrams mechanism is given by the dotted line. In this case its shape is 
closer to phase space.
\begin{figure}[h!]
\begin{center}
\rotatebox{270}{
\resizebox{8cm}{13cm}{
\includegraphics{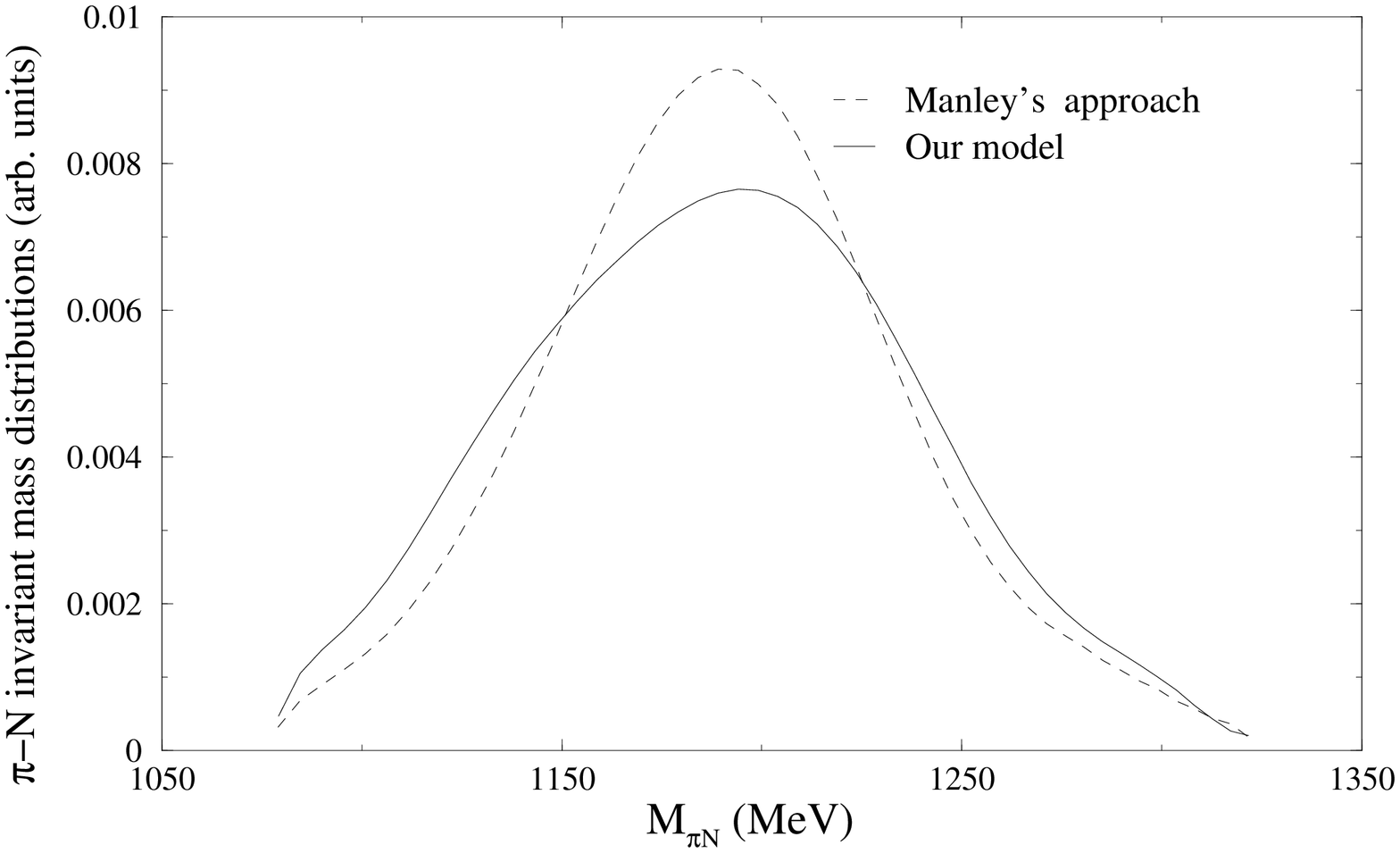}
}
}
\parbox[b]{14cm}{\caption{
Comparison of the $\pi-N$ invariant mass distributions as obtained 
in the two models.
Dashed line: Manley's approach distribution. Full curve: our model result. 
}}
\end{center}
\end{figure}
Both distributions are very different in  magnitude from
their corresponding $\Delta\pi$ and $N\epsilon$ in Manley's approach.
But once again the coherent sum of both gives a result close to Manley's as can be seen
in Fig. 11.

Our model is then able to reproduce the same phenomenology as Manley's approach
 without
the need of an explicit $R\to N\epsilon$ coupling. This $\epsilon$ or $\sigma$  meson
is generated dynamically in our model through the re-scattering of the
final pions in the appropriate channel. Even though our open- and
closed-diagrams contributions and their counterparts,  $\Delta\pi$ and
$N\epsilon$ channels, in Manley's  analysis  are individually quite different,
we have seen that the total result is very close in both models due to
interference.

\subsection{Extrapolation to low invariant masses}  
In this subsection we extrapolate our model to low invariant mass for the Roper. This region
for the Roper invariant mass is explored when studying two-pion production close to 
threshold in nucleon-nucleon
collisions. It is known that for the case where the two final pions are
in an isospin $I=0$ state
 a phenomenological two-pion s-wave coupling of the type
\begin{equation}
\label{lphen}
{\cal L}_{RN\pi\pi}= g_{RN\pi\pi}\Psi^{\dagger}_N(x)\Psi_R(x)\,\b{\pi}(x)\cdot
\b{\pi}(x)
\end{equation}
with $g_{RN\pi\pi}\simeq 1.6\cdot 10^{-2}$ MeV$^{-1}$ is able to explain
 the experimental
data~\cite{Alvarez-Ruso:1998mx}. The use of this  phenomenological term was
suggested by the Roper decay into nucleon plus two s-wave pions. The value
of the coupling constant in ref. ~\cite{Alvarez-Ruso:1998mx} was fitted to the
Roper decay width into $N(\pi\pi)_{S-wave}^{I=0}$ using for that the central
values given by the PDG~\cite{PDG:1996}.

Using our model we have evaluated the two-pion decay width of the Roper for
an  invariant mass of the latter given by  $M=1218$ MeV, just slightly
above the two-pion decay threshold. The contribution
coming from our open-diagrams is negligible as it should. The results
obtained with our closed-diagrams are collected in Table 2.

\begin{table}[h!]
\label{table:tab2}
\begin{center}
\nobreak
\begin{tabular}[t]{ |lr|   } 
\hline
& \\
 \hspace{1cm} Our Model & \\ 
& \\
\hline 
 & \\ 
$\bullet$\ Closed-diagrams & $1.1\cdot 10^{-3}$ \\  
& \\
\ \ - Delta alone & $3.2\cdot 10^{-4}$ \\
\ \ - No Delta    & $2.8\cdot 10^{-4}$ \\
 &  \\
\hline 
\end{tabular}
\parbox[b]{14cm}{
\caption{ Contribution of the closed-diagrams mechanisms to the two-pion
decay width of the Roper for invariant mass $M=1218$ MeV.
All numbers are in MeV. Notation as in Table 1.}}
\end{center}
\end{table}

At such a small invariant mass there is
almost no momentum dependence left on the amplitude. We can extract an
effective coupling constant to be compared to the one defined in Eq. (\ref{lphen}).
The value that we get is
\begin{equation}
g_{eff.}^{Our-model}=(6.5+i\ 2.0)\cdot 10^{-3} MeV^{-1}
\end{equation}
its imaginary part
coming from the physical cut where the intermediate nucleon and one of the pions
 appear on-shell.
  The main difference
is anyway in its module as
\begin{equation}
\frac{|g_{eff.}^{Our-model}|}{g_{RN\pi\pi}}=0.42
\end{equation}
This means that our model will under predict  the $L=0,I=0$ two-pion
production cross sections close to threshold. In contrast with our result
Manley's approach will give an effective coupling of
$g_{eff.}^{Manley}\simeq 1.43\cdot 10^{-2}$ MeV$^{-1}$ in agreement with
phenomenology. Our more fundamental model is able to reproduce Manley's
results at the Roper mass but it clearly lacks some extra contribution
at low invariant masses. It is clear that other mechanisms which would
contribute to the small amplitude at threshold are missing, which however would
not significantly contribute in the Roper region where the Delta intermediate
states give practically all the strength. Yet, it is still remarkable
that the mechanisms which we have evaluated provide the dominant contribution
over such a large span of energies.   

\section{Conclusions}
We have developed a model for the the two-pion decay of the Roper where we
take into account the re-scattering of the two final pions by means of the
  use of unitarized chiral theory. The aim was to see if one could explain
  at a more
fundamental level the need for the phenomenological decay channel
$R\to N\epsilon$ introduced by Manley and collaborators in their analyses.
Even though the two models have clear differences, we have shown that
we are able to reproduce the same phenomenology when
working at the Roper mass.

Our models seems nevertheless to fail at low invariant masses for the Roper,
indicating that we still lack some extra contributions.  Work in this direction
plus also work in the direction of obtaining microscopically a description of
the NN $\to$ N$N^*$ transition obtained in \cite{Hirenzaki:1996js} from the data
of \cite{Morsch:1992vj} would be a natural continuation of the present work,
complementing from the chiral symmetry perspective the work already done using
quark models in \cite{julia}.

\subsection*{Acknowledgments}
We would like to express our thanks to the $ECT^*$ Trento Center and the
organizer of the Roper Workshop, M. Soyeur, where the present subject was
discussed and clarified.
This work is 
partly supported by DGICYT contract number BFM2000-1326, 
the Junta de Castilla y
Leon under contract no. SA109/01
and the 
E.U. EURODAPHNE network contract no. ERBFMRX-CT98-0169.

\end{document}